%Paper: hep-th/9408145
%From: Alexander Laufer <alex@spock.physik.uni-konstanz.de>
%Date: Fri, 26 Aug 94 11:06:00 +0100
%Date (revised): Tue, 30 Aug 94 13:43:58 +0100

\magnification=1200
\abovedisplayskip=12pt plus 1pt minus 2pt
\overfullrule=0pt
\parindent=0pt
\advance\baselineskip by 3pt\
\font\titelf=cmbx10 scaled \magstep2
\font\gross=cmbx10 scaled \magstep1

\global\newcount\parnum
\global\newcount\glnum
\global\newcount\leerraum \leerraum=30
\def\newpar#1{\global\advance\parnum by 1
              \global\glnum = 0
              \bigbreak\bigbreak
               {\gross \the\parnum . \quad #1 }
              \bigskip}
%%
%  numbers of equations
%%
%
\def\gln{\global\advance\glnum by 1
           (\the\parnum .\the\glnum )}
\def\glnm{\global\advance\glnum by 1
          \eqno{(\the\parnum .\the\glnum )} }
\def\foono{\baselineskip=9pt  \sevenrm}
\def\abgl{\quad = \quad }
\def\tabgl{\quad &=\vphantom{\hbox to \leerraum pt {\hfill}} \quad }

\def\p{\; +\; }
\def\m{\; -\; }

\def\bruch#1#2{ \raise3pt \hbox {$ #1 $} / \lower3pt\hbox{$ #2 $}}

\def\cent{\centerline}
\def\ssk{\smallskip}
\def\bsk{\bigskip}
\def\msk{\medskip}
\def\vsk{\vskip}

\topinsert \vskip .5in \endinsert
\centerline{\titelf THE EXPONENTIAL MAP FOR THE }
\medskip
\centerline{\titelf   UNITARY GROUP SU(2,2) }
\vskip .5in
\cent{\bf A. O. Barut, $\;$ J. R. Zeni$\,$ {\dag }\footnote{*}{\foono
permanent address, Depto Ciencias Naturais, FUNREI, S\~ao Jo\~ao del
Rei, M.G., BRAZIL, 36.300.\hfil\break }
$\;$ and $\;$ A. Laufer$\,$\footnote{**}{\foono permanent address,
Physics Department, University of Konstanz, mailbox 5560 M678, 78434
KONSTANZ,\hfil\break GERMANY, e-mail
alex@spock.physik.uni-konstanz.de
\hfil\break}\footnote{\dag}{\foono work partially supported by
CAPES/BRAZIL and DAAD/GERMANY}}

\cent{Physics Department, University  of Colorado, Boulder, CO,
U.S.A.,
80.309-390}

\vsk 1in

{\bf Abstract} : In this article we extend our previous results for
the orthogonal group, $SO(2,4)$, to its homomorphic group $SU(2,2)$.
Here we present a closed, finite formula for the exponential of a
$4\times 4$ traceless matrix, which can be viewed as the generator
(Lie algebra elements) of the $SL(4,C)$ group. We apply this result
to the $SU(2,2)$ group,  which Lie algebra can be represented by the
Dirac matrices, and discuss how the exponential map for $SU(2,2)$ can
be written by means of the Dirac matrices.

\vfill\eject

\newpar{INTRODUCTION}

In a previous paper by these authors [Barut, Zeni and Laufer], we
obtained a closed, finite formula for the exponential map from the
Lie algebra into the defining representation of the orthogonal
groups, and in particular for the $SO_+(2,4)$ group. This result is a
generalization of the well known and important formulas for the
$SO(3)$ group [Barut, 80], and of the analogous result for the
Lorentz group, $SO_+(1,3)$ [Zeni and Rodrigues, 90].

\par
The present article deals with the exponential map from the Lie
algebra into the defining representation of $SU(2,2)$ group, which is
the covering group of the $SO_+(2,4)$ group. The result presented
here can be viewed as a generalization of the recent result to the
$SL(2,C)$ group [Zeni and Rodrigues, 92],
$$ e^F \abgl \cosh z \p \hat F \sinh z
\glnm $$
where $F = (e_i + {\bf i} \; b_i)\sigma_i$ is a complex vector
expanded in the Pauli matrices, the complex variable $z$ is such that
$z^2 = F^2$, and $\hat F = F/z$. We remark that the above result
contains the particular case of the $SU(2)$ group, when $e_i = 0$.
\ssk
The group $SU(2,2)$ and its homomorphic $SO_+(2,4)$ has several
applications in theoretical physics [Barut and Brittin]. For
instance, it is the largest group that leaves the Maxwell equations
invariant [Bateman]; from its subgroups, $SO(3)$ and $SO(2,1)$, we
can obtain the whole spectrum of the hydrogen atom as predicted by
Schr\"odinger theory [Barut, 72]; and more recently it has been used
in spin gauge theories as an attempt to generalize the minimal
coupling [Barut and McEwan], [Chisholm and Farwell], [Dehnen and
Ghaboussi].
\ssk
Also, the unitary groups play an important role in Quantum Mechanics
[Barut and Raczcka], and we can find in the literature several
articles
dealing with the parametrizations of these groups (see [Barnes and
Delbourgo] and references therein). For instance, [Bincer] presents a
parametrization for the exponential through a set of orthonormal
vectors that must be computed from the diagonal form of the matrix.
We remark that, if one uses our method to obtain the exponential
[Barut, Zeni and Laufer], we  need only to compute the eigenvalues of
the matrix, no further computations are involved, such as the
eigenvectors.
\par
We remark that, besides its application in group theory, the
exponential of a matrix has an important application in the solution
of a system of differential equations, as discussed in [Barut, Zeni
and Laufer], so the method developed in this present paper and in our
previous paper can be useful for that study. A comprehensive review
of methods to exponentiate an arbitrary matrix is given in [Moler and
van Loan] and references therein.
\ssk
The plan of the present paper is the following: in Section II we
obtain the exponential of a $4\times 4$ traceless matrix; in Section
III we discuss some particular cases of that exponential; in Section
IV we study the representation of the exponential in the Dirac
algebra, in particular, the cases when the generator is either the
sum of a vector and a axial vector, or a bivector; in Section V we
make our final comments.
\newpar{THE EXPONENTIAL OF A $4\times 4$ TRACELESS MATRIX}
The method presented previously by these authors [Barut, Zeni and
Laufer] can be generalized straightforward to exponentiate any given
matrix.  Basically, the algorithm presented was based on the
Hamilton-Cayley theorem and it emphasizes that the exponential must
be written by means of the eigenvalues. Also remarkable is the use of
the (square-root) discriminant related to the characteristic equation
of the matrix, as a multiplier to simplify the expressions of the
coefficients that appears in the recurrence relation. Finnaly, we
analyse the coefficients looking at each eigenvalue separately.
\ssk
We are going to apply the above algorithm to the generators of
$SL(4,C)$. We recall that the Lie algebra
 of $SL(n,C)$ is defined by,
$$ sl(n,C) \abgl \{ \; H \in C(n), \quad \hbox{such that,} \quad e^H
   \in SL(n,C) \} \glnm
$$
where $C(n)$ is the space of $n\times n$ complex matrices.
\par
Therefore the generators of $SL(n,C)$ are  traceless, since we have
$\det e^H = e^{\hbox{Tr} \; H}$.
\par
{}From now on we are restricting ourselves to $4 \times 4$ matrices.
\par
The characteristic equation of a $4 \times 4$  matrix is given by,
$$\eqalignno{
 \det \left( H \; -\; \lambda \, I \right) \tabgl \lambda^4 \m d_0 \,
  \lambda^3  \m a_0 \, \lambda^2 \m b_0 \, \lambda \m c_0   & \cr
  \tabgl (\lambda \; -\; w )\, ( \lambda \; -\; x )\, (\lambda \;
   -\; y )\,    (\lambda \; -\; z ) \abgl 0 & \gln  \cr}
$$
where $w$, $x$, $y$ and $z$ are the eigenvalues.
\par
The matrices representing the generators of $SL(4,C)$ are traceless,
so in this  case we have that the sum of the eigenvalues vanish,
$$
d_0 \abgl  w \p x \p y \p z \abgl \hbox{Tr} \; H \abgl 0 \glnm
$$
\ssk
{\bf The recurrence relations}
\ssk
The Cayley-Hamilton theorem says that a matrix satisfies a matrix
equation identical to its characteristic equation, therefore we can
write all higher powers of H in terms of the first powers
[Barut, Zeni and Laufer]
$$ H^{(4+i)} \abgl d_i \, H^3 \p a_i \, H^2 \p b_i \, H \p c_i \glnm
$$
So the series for $e^H$ becomes series in the coefficients of the
above equation.
Multiplying the recurrence relation, eq.(2.4), by $H$, and using the
Hamilton-Cayley theorem related to eq.(2.2), one obtains recurrence
relations for the coefficients, which holds for $i\geq 0$,
$$
   a_{i+1} \; = \; b_{i} \p d_{i} \, a_0, \qquad
   b_{i+1} \; = \; c_{i} \p d_{i} \, b_0, \qquad
   c_{i+1} \; = \; d_{i} \, c_0, \qquad
   d_{i+1} \; = \; a_{i}, \glnm
$$
{}From the above  relations we can also show that \quad $(i\geq 2)$,
$$
   a_{i+2}  \abgl a_{i}\; a_0 \p a_{i-1}\; b_0 \p a_{i-2}\; c_0
\glnm
$$
\par
In the next step, as outlined in [Barut, Zeni and Laufer], we
introduce the {\it square-root of the discriminant} of eq.(2.2),
indicated hereafter by m,
$$ m \abgl (w\m x)\; (w \m y) \; (w \m z) \; (x \m y) \; (x \m z) \;
    (y \m z) \glnm
$$
We write the first coefficients, $a_0$, $b_0$ and $c_0$ by means of
the eigenvalues, according to eq.(2.2), and from eq.(2.6) we find
that the general term for the coefficients $a_{i}$, multiplied
by $m$, is given by ($i \geq 0$)
$$ m \; a_{i}  \abgl t(w,y,z) \; x^{5+i} \p t(x,w,z) \;
y^{5+i} \p t(w,x,y) \; z^{5+i} \p t(y,x,z) \; w^{5+i}  \glnm
$$
where we made use of the {\it alternating}, $t(w,y,z)$, function of
three variables,
$$ t(w,y,z) \; = \; (w \m y) \; (y \m z) \; (z \m w)
\glnm $$
Now, based on eq.(2.8), the series for the coefficients can be summed
up easily, since we can write the other coefficients by means of
$a_i$, according to eq.(2.5). For instance,
$$ b_{i+1} \abgl a_{i-2} \; c_0 \p a_{i-1} \; b_0
\glnm $$
In order to write down the exponential, it is convenient to introduce
the {\it symmetric}, $s(w,y,z)$, function in three variables and the
{\it product} of the symmetric and alternating functions, indicated
hereafter by $st(w,y,z)$,
$$ s(w,y,z) \; = \; w \; y \p w \; z \p y \; z, \qquad
 \quad st(w,y,z) \; = \; s(w,y,z) \; t(w,y,z) \glnm
$$
\msk
{\bf The closed, finite formula for the exponential of
a $4 \times 4$ traceless matrix}
\ssk
$$ \eqalign{ m \; e^H & \abgl - \; w \; x \; y \; z \; \bigg(
t(w,y,z)
\; {e^x \over x} \p t(x,w,z) \; {e^y \over y} \p t(w,x,y) \; {e^z
\over
z} \p t(y,x,z) \; {e^w \over w} \bigg) \; 1 \cr \noalign{\vskip 10pt}
&
\p \bigg( st(w,y,z) \; e^x \p st(x,w,z) \; e^y \p st(w,x,y) \; e^z \p
st(y,x,z) \; e^w \bigg) \; H \cr \noalign{\vskip 10pt} & \p \bigg( x
\;
t(w,y,z) \; e^{x}  \p y \; t(x,w,z) \; e^{y} \p z \; t(w,x,y) \; e^z
\p
w \; t(y,x,z) \; e^w \bigg) \; H^2 \cr \noalign{\vskip 10pt} & \p
\bigg( t(w,y,z) \; e^x \p t(x,w,z) \; e^y \p  t(w,x,y) \; e^z \p
t(y,x,z) \; e^w \bigg) \; H^3 \cr \noalign{\vskip 7pt} }\glnm $$
\ssk
\newpar{SOME SPECIAL CASES OF THE EXPONENTIAL}
{\bf The case when \quad $b_0 = 0$.}
\par
Now we are going to see that the above formula for the exponential
simplifies considerably in the case when the characteristic equation
for $H$, eq.(2.2), has no term in the first power, i.e. $b_0 = 0$. In
this case the recurrence relations, eq.(2.4), for the even (odd)
powers involves only the even (odd) powers.
\par
If $b_0 = 0$ we have a quadratic equation in the square of the
eigenvalue of $H$, so we can set $w = - x$ and $z = - y$, and work
only
with two eigenvalues, $x$ and $y$. The Hamilton-Cayley theorem,
 related to eq.(2.2), becomes,
$$ H^4 \m (x^2 \, + \, y^2)\; H^2  \p x^2 \; y^2 \abgl 0 \glnm
$$
The square-root of the discriminant, $m$, eq.(2.7), reduces in this
case to,
$$  m \abgl \m 4 \; x \; y \; (x^2 \m y^2)^2 \glnm
$$
Therefore, the series for $e^H$, eq.(2.12), in the case when $b_0 = 0
$, is given by,
\ssk
$$\eqalign{ e^H & \abgl {x^2 \, \cosh y \m y^2 \, \cosh x \over x^2
\m
y^2} \; 1 \p {\cosh x \m \cosh y \over x^2 - y^2} \; H^2 \cr
\noalign{\vskip 10pt} & \p {x^3  \, \sinh y \m y^3 \, \sinh x \over x
\, y\, (x^2 \m y^2) } \; H \p {y \, \sinh x \m x \sinh y \over x\, y
\,
(x^2 \m y^2)}\; \, H^3 \cr } \glnm $$
\bsk
{\bf Further Simplifications: the case when \quad $H^2 \; = \; x^2\,
1$} \quad --- \quad In this case the square of the generator can be
identified with the square of one eigenvalue, say $x$, which is a
particular situation of the Hamilton-Cayley equation given above,
eq.(3.1). Examples includes the important cases when the generator is
either a vector (or axial vector), or a bivector from Dirac algebra.
This last case is just the Lorentz group.
\par
Therefore, substituting for $H^2 = x^2$ in eq.(3.3), we obtain,
$$ e^H \abgl \cosh x \p {\sinh x \over x} \; H \glnm
$$
\ssk
This is the formula given in [Zeni and Rodrigues, 92], for the
exponential of the generators of the Lorentz group.
We remark that [Zeni and Rodrigues, 92] proved in a very simple way
that every proper and ortochronous Lorentz transformation can be
written as the exponential of some generator.
\newpar{THE GROUP SU(2,2) AND THE DIRAC ALGEBRA}
\par
We recall that the Lie algebra of $SU(2,2)$ group is defined by
[Kihlberg et al],
$$ su(2,2) \abgl \left\{ H \in C(4), \quad \hbox{such that,}  \quad
H^{\dagger} \; \beta \; = \m \beta \; H
  \right\}\glnm $$
with $ \beta \; =\; \hbox{diag} (1,1,-1,-1) $.
\par
Also, the matrix algebra $C(4)$ is isomorphic to the Dirac algebra,
therefore the generators of $SU(2,2)$ can be represented by an
appropriate set of Dirac matrices.
\par
The {\it standard} representation for the Dirac matrices
is given by,
$$ \gamma_0 \; = \; \beta \; = \; \pmatrix{ I & 0 \cr 0 & -I \cr },
\qquad
   \gamma_i \; = \; \pmatrix{ 0 & \sigma_i \cr - \sigma_i & 0
   \cr }, \qquad \gamma_5 \; = \; \pmatrix{ 0 & - i \cr - i & 0 \cr }
   \glnm $$
where \quad $i \in [1,3]$ \quad and we set $\gamma_5 = \gamma_0 \,
\gamma_1 \, \gamma_2 \, \gamma_3$, so its square is {\it negative},
i.e., $\gamma_5\,^2 = - 1$, and it is {\it skew-hermitian}.
\par
A general element of the Lie algebra of $SU(2,2)$ can be written as
follows
$$ H \; = \; i \; V \p F \p \gamma_5 \; A \p i \; c \; \gamma_5 \; =
\;
i \; v^{\mu} \; \gamma_{\mu} \p F^{\mu \nu} \; \gamma_{\mu \nu} \p
\gamma_5 \; a^{\mu} \; \gamma_{\mu} \p i \; c \; \gamma_5 \glnm $$
where $\gamma_{\mu \nu} = \gamma_{\mu} \, \gamma_{\nu}$, \quad $\mu
\leq \nu$, \quad $\mu, \nu \in [0,3]$.
 The components $v^{\mu}$, $F^{\mu \nu}$, $a^{\mu}$, and $c$ are
{\it real} numbers.
\ssk
{\bf Vector $(H = i \, V$) or Axial Vector $(H = \gamma_5 \, A)$}:
\quad In this case, the square of the generator is a {\it real}
number,
$$ \eqalign{
\hbox{if},  \quad H \;  = \; i \; V \; = \; i\; v^{\mu} \;
\gamma{\mu},
\qquad & \Rightarrow \qquad H^2 \; = \; - \; V^2 \; = \; - \; v^{\mu}
\;
v_{\mu} \; = \; x^2 \cr \noalign{\vskip 5pt} \hbox{if,} \quad H  \; =
\; \gamma_5 \; A \; = \; \gamma_5 \; a^{\mu} \; \gamma_{\mu}, \qquad
& \Rightarrow \qquad H^2 \; = \; A^2 \; = \; a^{\mu} \; a_{\mu} \; =
\;
x^2 \cr } \glnm $$
In this case, the eigenvalues presents in eq.(3.1) are equal to each
other, i.e., we have $x^2 = y^2$.
\par
The previous formula for the exponential of $H$, eq.(3.4), holds in
every case, i.e., $V$ can be ligth-like ($V^2 =0$), space-like
($V^2 < 0$), or time-like ($V^2 > 0$).
\msk
{\bf Bivector $(H = F)$}: The square of a bivector in the Dirac
algebra can be {\it formally} identified with a complex number, which
is one of the eigenvalues, we say $x$. The other eigenvalue $y$ is
related to $x$ by complex conjugation. The imaginary unit is
represented by $\gamma_5$.
\par
Let us write the bivector as follows,
$$ F \abgl F^{\mu \nu} \; \gamma_{\mu \nu} \abgl (e^i \p \gamma_5 \;
b^i)\; \gamma_{i0} \glnm $$
In the latter form it is easy to compute the square of $F$ [Zeni and
Rodrigues, 92],
$$ F^2 \abgl \vec e \, ^2
\m \vec b \; ^2  \p 2 \; \gamma_5 \; \vec e {\bf .} \vec b \glnm
$$
{}From the above equation we can deduce the explicit form of the
Hamilton-Cayley theorem in this case,
$$ F^4 \m 2\,(\vec e \, ^2 \m \vec b \, ^2) \; F^2 \p 4 \, (\vec e
{\bf .} \vec b)^2 \p (\vec e \, ^2 \m \vec b \, ^2)^2 \abgl 0 \glnm
$$
\par
If we compare eq.(4.7) with eq.(3.1), we find that the eigenvalues
are given by,
$$ x^2 \; = \; \vec e \, ^2 \m \vec b \; ^2  \p 2 \; {\bf i} \;
  \vec e {\bf .} \vec b, \qquad \quad y^2 \; = \;  \vec e \, ^2 \m
  \vec b \; ^2 \m 2 \; {\bf i} \;  \vec e {\bf .} \vec b \glnm
$$
Observe that the expression for $x$ is the same expressions as that
for $F^2$. We have only to replace the imaginary unit for
$\gamma_5$. Therefore, eq.(3.4) applies for the exponential of a
bivector.
\msk
{\bf The Sum of a Vector and an Axial Vector \quad $(H \, = \, W \, =
\, i \, V + \gamma_5 \, A)$}: \quad In this case we are going to show
that the fourth power of the generator can be written by means of the
second power and the identity, so eq.(3.3) applies for the
exponential
of a {\it V-A} generator.  Also, we obtain an explicit expression for
the second and third power by means of $V$ and $A$, eq.(4.16), which
can simplify further computations with the exponential, eq.(3.3).
{}From now on, we are indicating $W = H$ for the generator, so we
have,
$$
 W \abgl  i \, V \; + \; \gamma_5 \, A \abgl (i\, V^{\mu} \; + \;
 \gamma_5 \, A^{\mu}) \gamma_{\mu} \glnm
$$
Computing the second and fourth power of W we find,
$$
 \eqalignno{
  W^2 & \abgl A^2 \m  V^2 \p  i \; \gamma_5 \, ( A \, V - V \, A) &
  \gln \cr \noalign{\vskip 7pt} W^4 & \abgl 2\, (A^2 \m V^2) \, W^2
 \p 4 \, (A\bullet V)^2 \m (A^2 \p V^2)^2 & \gln \cr }
$$
Therefore, a V-A generator satisfies eq.(3.1) and it can be
exponentiated as in eq.(3.3).
\par
Observe that if, $\, V = A$, i.e., $W \, = \, (1 +
\gamma_5)V$, it implies $\; W^2 \, = \, 0$, and $e^W \, = \, 1 + W$.
\ssk
We introduced $\; A\bullet V$ as the {\it inner product} in the Dirac
algebra (also $V^2 \, = \, V \bullet V$),
$$
 V \bullet A \abgl 1/2 \, ( V \, A \p A \, V) \abgl v^{\mu} \,
a_{\mu}
$$
\ssk
We also remark that \qquad $ (A\, V - V \, A)^2  \abgl 4 \, (A\bullet
V)^2 \; - \; 4 \, A^2 \, V^2 \quad {\buildrel \rm def \over =} \quad
\Delta$.
\ssk
If we compare eq.(4.11) with eq.(3.1) we see that the eigenvalues are
given by,
$$ x^2 \; = \; A^2 \m V^2 \p  \sqrt{\;\Delta}, \qquad \quad
 y^2 \; = \; A^2 \m V^2 \m \sqrt{ \;\Delta}
\glnm $$
\ssk
To get a convenient expression for the third power of $W$ we
introduce a new element, $W^*$, defined by,
$$
 W^* \abgl  A \p  i \; \gamma_5 \; V \glnm
$$
The products $W \; W^*$ and $W^* \; W$ will be called here {\it
bicomplex numbers}, i.e., we have now two imaginary {\it commutative}
units, $\gamma_5$ and $i$.  Moreover, the above products are related
to each other through complex conjugation respective to $\gamma_5$,
i.e., we have,
$$
 \eqalign{ u \quad & {\buildrel \rm def \over =} \quad  W \; W^*
\abgl  2 \, i \; A\bullet V \; + \; (A^2 \p V^2) \; \gamma_5 \cr
\noalign{\vskip 7pt} \bar u  \quad & {\buildrel \rm def \over =}
\quad
W^* \; W \abgl   2 \, i \; A\bullet V \m (A^2 \p  V^2) \; \gamma_5
\cr } \glnm
$$
It is remarkable that the product $u \; \bar u = \bar u \; u$ is a
real number, which is just the {\it determinant} of $H = W$ in the
matrix representation (cf. eq.(4.11) above),
$$
 u \; \bar u \abgl \bar u \; u \abgl \m 4 \, (A\bullet V)^2 \p (A^2
\p V^2)^2 \glnm
$$
Based on eq.(4.15),  we obtain the inverse of $W$ as
(when there is an inverse, i.e., $u \; \bar u \neq 0$),
$$
 W^{-1} \abgl  W^* \; u^{-1} \abgl {W^* \; {\bar u} \over u \;
 \bar u}  \glnm
$$
To verify that the above expression just defines the {\it bilateral}
inverse, we call attention to the fact that $\gamma_5$ anticommutes
with $W$, so we have $ {\bar u}\, W  \; = \; W \, u $.
Now considering that $ W^3 \; = \; W^4 \; W^{-1}$, it follows from
eq.(4.11) and eq.(4.16) that the third power of $W$ is given by (if,
$\; \bar u \, u \neq 0$),
$$
 W^3 \abgl 2 \, (A^2 \m V^2) \; W \m W^* \; \bar u \glnm
$$
which is easily expressed by means of the Dirac matrices, through
eq.(4.9) and eq.(4.13).
\par

\newpar{CONCLUSIONS}

In this article we presented a finite, closed formula for the
exponential of a $4\times 4$ traceless matrix, eq.(2.12). It can be
viewed as the exponential of a generator of the $SL(4,C)$ group,
which
includes the $SU(2,2)$ group as a subgroup. Our approach to get the
exponential is based on our previous work [Barut, Zeni and Laufer].

Eq.(2.12) is a generalization of the exponential for generators of
the
$SL(2,C)$ group presented in [Zeni and Rodrigues, 92].

The finite formula for the exponential, eq.(2.12), involves only the
computations of the eigenvalues and the first three powers of the
matrix, no further computations (e.g., eigenvectors) are needed to
obtain the exponential (cf. [Moler and van Loan]).

We have also presented some special cases of this exponential,
eq.(3.3)
and eq.(3.4). They include the important cases when the generator of
the $SU(2,2)$ group is identified either with a bivector or the sum
of
a vector and a axial vector, as discussed in Section IV.  For both
cases, we gave explicit expressions for the eigenvalues of the
generators, eq.(4.8) and eq.(4.12), as derived from Dirac algebra .
Moreover, in the case of a V-A generator, we obtained a simple
expression for the third power of the generator by means of $V$ and
$A$, eq.(4.17), which is needed to exponentiate the generator (see
eq.(3.3)).

\msk
{\bf ACKNOWLEDGMENTS}
\ssk

The authors are thankful to Prof. W.A. Rodrigues Jr., P. Lounesto,
and H. Dehnen for several discussions on the subject. J. R. Zeni and
A. Laufer are thankful to Prof. A. O. Barut for his kind hospitality
in Boulder. A. Laufer and J. R. Zeni are also thankful to
DAAD/GERMANY and CAPES/BRAZIL, respectively, for the fellowships that
supported their stay in Boulder.

\ssk
{\bf REFERENCES}
\ssk

K.J. BARNES and R. DELBOURGO (1972) J. Phys. {\bf A 5}, 1043-1053.

A.O. BARUT (1964) Phys. Rev., {\bf 135}, B839-B842.

A.O. BARUT (1980) {\it Electrodynamics and Classical Theory of Fields
and Particles}, DOVER.

A.O. BARUT (1972) {\it Dynamical Groups and Generalized Symmetries in
Quantum Theory}, Univ. of Canterbury Press, Christchurch, New
Zealand.

A.O. BARUT and W.E. BRITTIN (editors, 1971) in {\it Lectures on
Theoretical Physics: De Sitter and Conformal Groups}, vol. XIII,
Colorado Univ. Press, Boulder.

A.O. BARUT and J. McEWAN (1984) Phys. Lett., {\bf 135}, 172-174.

A.O. BARUT and R. RACZKA (1986) {\it Theory of Group Representation
and
Applications}, second ed., World Scientific.

A.O. BARUT, J.R. ZENI and A. LAUFER (1994) ``The Exponential Map for
the Conformal Group O(2,4),'' in press, J. of Phys. A.

H. BATEMAN (1910) Proc. London Math. Soc., {\bf 8}, 288.

A.M. BINCER (1990) J. Math. Phys., {\bf 31}, 563-567.

J.S.R. CHISHOLM and R.S. FARWELL (1989) J. of Phys. A, {\bf 22},
1059-1071.

H. DEHNEN and F. GHABOUSSI (1986) Phys. Rev. D, {\bf 33}, 2205-2211.

A. KILHBERG, V.F. M\"ULLER and F. HALBWACHS (1966) Commun. Math
Phys.,
{\bf 3}, 194-217.

C. MOLER and C. VAN LOAN (1978) SIAM Review, {\bf 20}, 801-836.

J.R. ZENI and W.A. RODRIGUES (1990) Hadronic J., {\bf 13}, 317-323.

J.R. ZENI and W.A. RODRIGUES (1992) Int. J. Mod. Phys., {\bf A7},
1793-1817.

\vfill\end